\newcommand{\PRD}[3]{{\it Phys. Rev.} {\bf D{#1}} ({#2}), {#3}}

\newcommand{\NPB}[3]{{\it Nucl. Phys.} {\bf B{#1}} ({#2}), {#3}}

\newcommand{\PLB}[3]{{\it Phys. Lett. } {\bf B{#1}} ({#2}), {#3}}

\newcommand{\PTP}[3]{{\it Prog. Theor. Phys.} {\bf {#1}} ({#2}), {#3}}

\newcommand{\JHEP}[3]{{\it J. High Energy Physics.} {\bf {#1}} ({#2}), {#3}}
\newcommand{\JP}[3]{{\it J. of Phys.} {\bf {#1}} ({#2}), {#3}}
\newcommand{\ANN}[3]{{\it Ann. of Phys.} {\bf {#1}} ({#2}), {#3}}

\documentclass[12pt]{article}
\usepackage{graphicx,amsmath,amssymb,amscd}
\usepackage{bm}

\begin{document}
\baselineskip=18pt
\begin{titlepage}
\begin{flushright}
KOBE-TH-11-06
\end{flushright}
\vspace{1cm}
\begin{center}{\Large\bf 
Phase Structure of Gauge Theories on an Interval
}
\end{center}
\vspace{0.5cm}
\begin{center}
Yukihiro Fujimoto$^{(a)}$
\footnote{E-mail: 093s121s@stu.kobe-u.ac.jp},
Tomoaki Nagasawa$^{(b)}$
\footnote{E-mail: nagasawa@gt.tomakomai-ct.ac.jp},
Satoshi Ohya$^{(c)}$
\footnote{E-mail: ohya@mri.ernet.in}\\
and Makoto Sakamoto$^{(a)}$
\footnote{E-mail: dragon@kobe-u.ac.jp}
\end{center}
\vspace{0.2cm}
\begin{center}
${}^{(a)}$ {\it Department of Physics, Kobe University,
		1-1 Rokkodai, Nada, Kobe 657-8501, Japan}
\\[0.2cm]
${}^{(b)}$ {\it Tomakomai National College of Technology,
		443 Nishikioka, Tomakomai 059-1275, Japan}
\\[0.2cm]
${}^{(c)}$ {\it Harish-Chandra Research Institute,
        Chhatnag Road, Jhusi, Allahabad 211 019, India}
\end{center}







\vspace{1cm}
\begin{abstract}
We discuss gauge symmetry breaking in a general framework of 
gauge theories on an interval.
We first derive a possible set of boundary conditions 
for a scalar field, which are compatible with several consistency 
requirements.
It is shown that with these boundary conditions the scalar field 
can acquire a nontrivial vacuum expectation value even if the scalar
mass square is positive.
Any nonvanishing vacuum expectation value cannot be a constant but, 
in general, depends on the extra dimensional coordinate of the interval.
The phase diagram of broken/unbroken gauge symmetry possesses 
a rich structure in the parameter space of the length of the interval, 
the scalar mass and the boundary conditions. 
We also discuss 4d chiral fermions and fermion mass hierarchies 
in our gauge symmetry breaking scenario.
\end{abstract}
\end{titlepage}

\newpage

\section{Introduction}

Despite a great success of the standard model (SM), the Higgs sector still remains a mystery.
A full understanding of the role of the Higgs scalar is a key ingredient in constructing realistic models beyond the SM.
The Higgs scalar of the SM plays two roles: 
A nonvanishing vacuum expectation value (VEV) of the Higgs field breaks the electroweak gauge symmetry and produces the fermion masses.
To break the gauge symmetry, a negative mass square of the Higgs boson is required.
Since fermion mass terms are kinematically forbidden in the SM, all the fermions (probably except for neutrinos) acquire their masses through the Yukawa couplings to the Higgs scalar.
Therefore, the fermion mass hierarchy originates in the Yukawa coupling one.
The purpose of this paper is to discuss gauge symmetry breaking,
the chiral structure of fermions and Yukawa coupling hierarchies
from a viewpoint of gauge theories on an interval.

	An attractive mechanism for generating the Yukawa coupling hierarchy 
has been proposed by Arkani-Hamed and Schmaltz (AS) \cite{AS2000}.
It is naturally realized by localizing the SM fermions at different points 
in one (more) extra dimension(s), which can be done by coupling 
five-dimensional fermions to a scalar field with a kink background
configuration \cite{RS1983,Akama1983}.
The AS mechanism has been extended by further requiring that 
the Higgs scalar VEV is localized in one extra dimension \cite{DS2000,KT2001}.
The resulting fermion masses are determined by exponentially suppressed 
overlaps of their wavefunctions and become automatically hierarchical.
In order for the above scenario to work, it is important to realize 
a mechanism to generate an extra dimensional coordinate dependent VEV 
of a scalar field as a ground state configuration.
A mechanism to break translational invariance by a scalar VEV has 
been proposed in Refs. \cite{STT1999a}-\cite{HMOS2001}.
The idea is to impose nontrivial boundary conditions (BC's) 
incompatible with a nonvanishing constant configuration of a scalar field.
The mechanism has been used to break supersymmetry \cite{STT1999b,STT2000} 
and extended to higher extra dimensions \cite{MST2001,ST2002}.
A similar mechanism to produce a nontrivial profile of a scalar field 
has been found in orbifold models that orbifold BC's can be chosen
to forbid a nonvanishing constant scalar configuration \cite{GGH2001}.
Another mechanism is to prepare a brane localized potential 
in addition to a bulk potential in such a way that a minimum 
configuration of the bulk potential conflicts with that of 
the brane one \cite{KT2001}.
This will force a scalar VEV to depend on the extra dimensional coordinate.
The Higgs scalar can also develop a nontrivial profile along the extra
dimension by introducing a coupling to another scalar of the 
localizer \cite{DS2000,KT2001}.


In this paper, we try to find an answer to naturally explain
the physics of the Higgs sector in a framework of gauge theories
on an interval.
We assume that all fields live on the bulk \cite{ACD2001} with
no brane/boundary localized term, and that any model in our setting
is specified by a bulk Lagrangian and BC's for fields.
Since BC's at the boundaries of the interval are crucially 
important in our framework, we first derive a general class
of possible BC's for a scalar field on an interval, which
are compatible with several consistency requirements.
Those BC's of a scalar are wider than the commonly used orbifold 
BC's \cite{HN2001}-\cite{HM2002} and characterized by two parameters.
In this general setting of the BC's, we find 
that the scalar field can acquire a nonvanishing vacuum 
expectation value even if the scalar mass square is positive 
with no boundary localized terms.
Furthermore, the VEV turns out to inevitably depend on the extra 
dimensional coordinate of the interval.
As an illustrative example, we consider a scalar QED on an interval 
and show that the phase diagram of the broken/unbroken gauge symmetry 
has a rich structure, which complicatedly depends on the length of the interval, the scalar mass square and the parameters specifying the BC's.
	
We continue to derive consistent BC's for a fermion on an interval 
and find that only the four types of BC's are allowed.
The type ($++$) or ($--$) BC leads to a 4d massless chiral fermion 
even if the fermion has a bulk mass, while the type ($+-$) or ($-+$) 
BC produces no massless chiral fermions.
Thus, only fermions which obey the type($\pm$$\pm$) BC's survive 
at low energies as 4d massless chiral fermions.
All other Kaluza-Klein modes will be decoupled from the low energy 
spectrum with masses of the order of $L^{-1}$, which is the inverse 
of the length of the interval.
An important observation is that the profile of a chiral zero mode 
is exponentially localized at one of the boundaries of the interval.
Since chiral fermions could acquire their masses through the Yukawa 
couplings to the Higgs scalar, it will not be surprising that fermions 
get hierarchically different masses through the Yukawa interactions 
because of exponentially localized profiles of chiral zero modes 
as well as the Higgs VEV with a nontrivial extra dimensional 
coordinate dependence.
Thus, our setting of gauge theories on an interval may be regarded 
as an explicit realization of the scenario given in 
Refs. \cite{DS2000,KT2001}.
The above observations strongly suggest that a mystery of the Higgs 
sector in the SM can be naturally solved in a framework of gauge 
theories on an interval.
Our considerations will mostly be restricted to 
abelian gauge theories in this paper.
However, our mechanism to break gauge symmetries can work for 
nonabelian gauge theories, as well.
	
This paper is organized as follows.
In the next section, we determine a general consistent set of BC's 
for a scalar field on an interval.
In Section 3, we investigate a scalar QED on an interval, 
as a demonstration of our symmetry breaking mechanism, and show that 
the scalar can acquire a nontrivial VEV even if the mass square 
of the scalar is positive.
Furthermore, we clarify a rich phase structure of the model.
In Section 4, we consider a fermion on an interval and show 
a possible class of BC's, in which some of the BC's lead to 
4d massless chiral fermions.
The Section 5 is devoted to conclusions.

\section{Consistent BC's of a Scalar Field}
%
In this section, we investigate a complex scalar on an interval 
and clarify a class of general consistent BC's for the scalar field.
Since the BC's for scalars play a crucial role in our mechanism 
to break gauge symmetries, we shall discuss the consistency of 
the allowed BC's from various different points of view.
To this end, let us consider a complex scalar field $\Phi(x,y)$ 
on an interval with an action
\begin{align}
	S=\int d^4 x\int^{L}_{0}dy \Bigl\{ \Phi^{\ast}\partial^{\mu}\partial_{\mu}\Phi+\Phi^{\ast}\partial_{y}^2\Phi -V(|\Phi|^2)\Bigl\},\label{eq:1}
\end{align}
where $x^{\mu} (\mu=0,1,2,3)$ denotes the coordinate of the four-dimensional Minkowski spacetime and $y$ is the coordinate of the extra dimension with $0\leq y \leq L$.
Here, the 5d metric is chosen as $\eta_{KN}=\textrm{diag}(-,+,+,+,+)$.
	
In one-dimensional quantum mechanics, the most general BC's of 
a wavefunction are known to be characterized by $U(2)$ parameters 
at a boundary or point singularity. \cite{RS1975}-\cite{CFT2001}
If the probability current is required to vanish at a boundary, 
the $U(2)$ parameters reduce to a subfamily of $U(2)$ at each boundary.
Since an interval has two boundaries, at which the probability current 
has to vanish in order to preserve the probability conservation, 
the allowed boundary conditions on an interval are found to be 
given by the Robin boundary condition\!\!
\footnote{
In order to concentrate on the extra dimensional coordinate $y$, 
we will omit the $x^{\mu}$ dependence unless otherwise stated.
}
%
\begin{align}
	\Phi(0)+L_{+}\partial_{y}\Phi(0)=0,\nonumber\\
	\Phi(L)-L_{-}\partial_{y}\Phi(L)=0, \label{eq:2}
\end{align}
where $L_{\pm}$ are arbitrary real constants of mass dimension $-1$.
	
The boundary conditions (\ref{eq:2}) can also be obtained from 
the hermiticity requirement of the action, which is necessary 
to ensure the unitarity of the theory.
The condition $S^{\dagger}=S$ immediately leads to
\begin{align}
	j(y)\equiv -i\Bigl(\Phi^{\ast}(y)\partial_{y}\Phi(y)-(\partial_{y}\Phi^{\ast}(y))\Phi(y)\Bigl)=0 \hspace{1.3em}{\rm at}\hspace{0.5em} y=0,L, \label{eq:3}
\end{align}
where we have assumed that the field and its derivatives become zero 
at $|x^{\mu}|\rightarrow \infty$, as usual.
The equations (\ref{eq:3}) can be solved by rewriting it as \cite{NOSSS2009}
\begin{align}
	|\Phi -iL_{0}\partial_{y}\Phi|^2=|\Phi +iL_{0}\partial_{y}\Phi |^2 \hspace{1.3em}{\rm at}\hspace{0.5em} y=0,L, 
\end{align}
where $L_{0}$ is an arbitrary nonzero real constant of mass dimension $-1$.
The above equations imply that $\Phi-iL_{0}\partial_{y}\Phi$ should 
be proportional to $\Phi+iL_{0}\partial_{y}\Phi$ at $y=0,L$ and 
the proportional constants are given by phase factors.
Thus, we find that 
$\Phi-iL_{0}\partial_{y}\Phi =e^{i\theta_{0,L}}(\Phi+iL_{0}\partial_{y}\Phi)$
at $y=0,L$ and obtain Eq.(\ref{eq:2}) with the identification 
$L_{+}=L_{0}\cot\frac{\theta_{0}}{2}$ and 
$L_{-}=-L_{0}\cot\frac{\theta_{L}}{2}$.
	
Another way to derive the BC's (\ref{eq:2}) is to impose the 
conservation of a $U(1)$ charge.
The action (\ref{eq:1}) is invariant under the global $U(1)$ 
transformation: $\Phi\rightarrow e^{i\alpha}\Phi$, and the $U(1)$ 
current $j_{N}\equiv -i\bigl(\Phi^{\ast}\partial_{N}\Phi 
-(\partial_{N}\Phi^{\ast})\Phi\bigr)$ $(N=0,1,2,3,y)$ will be 
conserved, i.e. $\partial^{N}j_{N}=0$.
However, this does not necessarily assure the conservation of the 
$U(1)$ charge
\begin{align}
	Q\equiv \int d^3 \boldsymbol{x}\int^{L}_{0}dy  \hspace{0.3em}   
	j^{0}(x,y),
\end{align}
because
\begin{align}
	\frac{dQ}{dt}=-\int d^3 \boldsymbol{x}\int^{L}_{0}dy \hspace{0.3em}\partial_{y}j_{y}=-\int d^3 \boldsymbol{x}\Bigl(j_{y}(x,L)-j_{y}(x,0)\Bigl).
\end{align}
Thus, the $U(1)$ charge conservation can be achieved only 
when the extra dimensional component of the current $j_{y}$ 
vanishes at the boundaries $y=0$ and $L$.
This is identical to the conditions (\ref{eq:3}), so that 
the previous argument shows that the BC's (\ref{eq:2}) guarantee 
the conservation of the $U(1)$ charge.
We should emphasize that the conservation of the global $U(1)$ charge 
is very important, otherwise the model cannot be extended to 
a local gauge invariant theory on an interval.
	
As was proposed in Ref. \cite{CGMPT2004}, consistent BC's will be 
obtained from the action principle.
For any infinitesimal variations of $\Phi$, the requirement 
$\delta S=0$ leads to the bulk field equation for $\Phi$ 
(or $\Phi^{\ast}$), together with the boundary equations
\begin{align}
	\Phi^{\ast}\partial_{y}\delta\Phi -(\partial_{y}\Phi^{\ast})\delta\Phi=0 \hspace{1.3em}{\rm at}\hspace{0.5em} y=0,L.\label{eq:7}
\end{align}
Since $\Phi^{\ast}$ and $\delta \Phi$ can be independent each other, 
the above conditions seem to be more restrictive than Eq.(\ref{eq:3}).
This is not, however, the case.
The equation (\ref{eq:7}) is found to lead to the same BC's (\ref{eq:2}).
Indeed, it is easy to see that Eq.(\ref{eq:7}) is satisfied if 
both $\Phi$ and $\delta \Phi$ obey the BC's (\ref{eq:2}).
	
Before closing this section, we should make a comment on the form of 
the action (\ref{eq:1}).
We could start with the action
\begin{align}
	S'=\int d^4 x\int^{L}_{0}dy \Bigl\{-\partial_{\mu}\Phi^{\ast}\partial^{\mu}\Phi -\partial_{y}\Phi^{\ast}\partial_{y}\Phi -V(|\Phi|^2)\Bigl\}, \label{eq:8}
\end{align}
instead of Eq.(\ref{eq:1}).
Then, the action principle will lead to the BC's
\begin{align}
	(\partial_{y}\Phi^{\ast})\delta\Phi=0  \hspace{1.3em}{\rm at}\hspace{0.5em} y=0,L, \label{eq:9}
\end{align}
which would require the BC's\!\!
\footnote{
The equation (\ref{eq:9}) would allow the BC $\Phi ={\rm const}$ 
at $y=0,L$ \cite{CGMPT2004}.
Any nonvanishing constant value of $\Phi$ at the boundaries, however, 
turns out to be inconsistent with gauge invariance when the scalar field 
is coupled to a gauge field.
This is a signal of the violation of unitarity \cite{SU2007}.
Thus, we do not consider this possibility in this paper, although 
there is an argument that such a boundary condition probably gives 
a consistent unitary theory \cite{NO2011}.
}
%
\begin{eqnarray}
	\partial_{y}\Phi=0\hspace{0.8em}{\rm or}\hspace{0.8em}\Phi=0 \hspace{1.7em}{\rm at}\hspace{0.5em} y=0,L.
\end{eqnarray}
These are special cases (i.e. $L_{\pm}=\infty$ or $L_{\pm}=0$) of 
the BC's (\ref{eq:2}).
In order to obtain the BC's (\ref{eq:2}), we may add the following boundary terms
\begin{align}
\int d^4 x\int^{L}_{0}dy\Bigl\{-\Bigl(-\frac{1}{L_{+}}\delta(y)-\frac{1}{L_{-}}\delta(y-L)\Bigl)|\Phi(y)|^2\Bigl\},\label{eq:11}
\end{align}
to the above action (\ref{eq:8}). Then, the action principle is found to reproduce the BC's (\ref{eq:2}).
In this point of view, physical meanings of the parameters $L_{\pm}$ become clear.
For $L_{\pm}>0$ $(L_{\pm}<0)$, the terms in Eq.(\ref{eq:11}) correspond to attractive (repulsive) 
$\delta$-function potentials at the boundaries with the couplings $1/L_{\pm}$. 
This interpretation will be helpful to understand the phase structure of gauge symmetry breaking, as
we will discuss later.

\section{Phase Structure of Scalar QED on an Interval}
%
In this section, we investigate a scalar QED on an interval, 
as an illustrative example of our symmetry breaking mechanism, 
and show that our mechanism possesses notable properties different 
from the standard Higgs mechanism.
As was noted in the Introduction, we assume that all the fields live
in the bulk without any brane/boundary localized term, and that
the scalar field is allowed to obey a class of the general BC's
(\ref{eq:2}).

\subsection{Scalar QED Action and BC's}
%
The action we consider is
%
\begin{align}
		S=\int d^4 x\int^{L}_{0}dy\Bigl\{-\frac{1}{4}F_{KN}F^{KN}+\Phi^{\ast}D_{\mu}D^{\mu}\Phi+\Phi^{\ast}D_{y}^2\Phi -M^2\Phi^{\ast}\Phi-\frac{\lambda}{4}(\Phi^{\ast}\Phi)^2\Bigl\},\label{eq:3-1}
\end{align}
where $F_{KN}=\partial_{K}A_{N}-\partial_{N}A_{K}$ and 
\begin{align}
		D_{N}\Phi =(\partial_{N}-ieA_N )\Phi, \hspace{1.3em}N=0,1,2,3,y.
\end{align}
%
%
As discussed in the previous section, we take the following BC's 
for the scalar:
\begin{align}
		\Phi(0)+L_{+}\partial_{y}\Phi(0)=\Phi(L)-L_{-}\partial_{y}\Phi(L)=0.\label{eq:3-3}
\end{align}
For the gauge fields $A_{M}$, we choose the BC's to be of the form
\begin{align}
		\partial_{y}A_{\mu}(0)=\partial_{y}A_{\mu}(L)=0,\label{eq:3-4}\\
		A_{y}(0)=A_{y}(L)=0.\label{eq:3-5}
\end{align}
Since we are interested in gauge symmetry breaking through 
a nontrivial VEV of $\Phi$, the BC's for the gauge fields have 
to be chosen not to break the 4d gauge symmetry explicitly.
In fact, the BC's (\ref{eq:3-4}) allow a massless zero mode of 
the 4d gauge field, as they should be.

It is important to verify that the BC's (\ref{eq:3-4}) and (\ref{eq:3-5}) 
are consistent with the 5d gauge transformations:
\begin{align}
		\Phi(x,y)&\rightarrow \Phi'(x,y)=e^{ie\Lambda(x,y)}\Phi(x,y),\label{eq:3-6}\\
		A_{\mu}(x,y)&\rightarrow A'_{\mu}(x,y)=A_{\mu}(x,y)+\partial_{\mu}\Lambda(x,y),\label{eq:3-7}\\
		A_{y}(x,y)&\rightarrow A'_{y}(x,y)=A_{y}(x,y)+\partial_{y}\Lambda(x,y)\label{eq:3-8}.
\end{align}
It follows from the transformation (\ref{eq:3-7}) that the gauge 
parameter $\Lambda(x,y)$ should obey the same BC's as $A_{\mu}(x,y)$, i.e.
\begin{align}
		\partial_{y}\Lambda(x,0)=\partial_{y}\Lambda(x,L)=0,\label{eq:3-9}
\end{align}
which is consistent with the transformation (\ref{eq:3-8}) and 
the BC's (\ref{eq:3-5}) for $A_{y}$.
Note that the compatibility between the BC's (\ref{eq:3-4}) and 
(\ref{eq:3-5}) can also be shown from a viewpoint of quantum mechanical
supersymmetry. \cite{NOSSS2009,LNSS2005}-\cite{LNOSS2008b}

The consistency of the BC's (\ref{eq:3-3}) for $\Phi$ with the gauge
transformation (\ref{eq:3-6}) requires that $\Phi'$ given in 
Eq.(\ref{eq:3-6}) should obey the same BC's as the original field $\Phi$.
This can be verified as follows:
\begin{align}
		\Phi'(y)\pm L_{\pm}\partial_{y}\Phi'(y)
		=e^{ie\Lambda(y)}\Bigl\{\Bigl(\Phi(y)\pm L_{\pm}\partial_{y}\Phi(y)\Bigl)\pm ie(\partial_{y}\Lambda(y))L_{\pm}\Phi(y)\Bigl\}.
\end{align}
It is now easy to see that $\Phi(y)$ and $\Phi'(y)$ satisfy 
the same BC's (\ref{eq:3-3}) with the conditions (\ref{eq:3-9}).
		
It should be further noticed that the BC's (\ref{eq:3-3}), (\ref{eq:3-4}) 
and (\ref{eq:3-5}) satisfy all the requirements discussed in the previous
section.
The action (\ref{eq:3-1}) is hermitian and the $U(1)$ charge is 
conserved (if the gauge symmetry is unbroken).
The boundary equations derived from the action principle are satisfied 
for the BC's chosen here.

\subsection{Phase Structure}
%
In this subsection, we would like to determine whether or not 
the scalar field $\Phi$ can acquire a nonzero VEV.
In order to find the vacuum configuration, one might try to minimize 
the potential $V=M^2|\Phi|^2+\frac{\lambda}{4}|\Phi|^4$.
This is, however, wrong in the present model.
It turns out that the vacuum configuration is given by solving 
the minimization problem of the functional
\begin{align}
		{\cal E}[\Phi]\equiv \int^{L}_{0}dy\Bigl\{\Phi^{\ast}(-\partial_{y}^2)\Phi +M^2|\Phi|^2 +\frac{\lambda}{4}|\Phi|^4\Bigl\}.
\end{align}
The important point is to incorporate the kinetic term along 
the extra dimension into the effective 4d potential ${\cal E}[\Phi]$, 
because the minimum configuration of $\Phi$ can have the $y$ dependence, 
as we will see below.
		
In the following, we will ignore the $x^{\mu}$ dependence in $\Phi$, 
which is irrelevant in our analysis since we assume the translational
invariance of the 4d Minkowski spacetime.
Since we are interested in the gauge symmetry breaking, we would like 
to know whether or not the vacuum configuration $\langle\Phi(y)\rangle$ 
is nonvanishing.
To this end, it is convenient to introduce the eigenfunctions $f_{n}(y)$ 
of the eigenvalue equation
\begin{align}
		-\partial_{y}^2 f_{n}(y)=E_{n}f_{n}(y),\hspace{1.3em}n=0,1,2,\cdots,\label{eq:3-12}
\end{align}
with the BC's
\begin{align}
		f_{n}(0)+L_{+}\partial_{y}f_{n}(0)=f_{n}(L)-L_{-}\partial_{y}f_{n}(L)=0.\label{eq:3-13}
\end{align}
In terms of the orthonormal eigenfunctions $f_{n}$, the field $\Phi$ 
can be expanded as
\begin{align}
		\Phi(y)=\sum^{\infty}_{n=0}\phi_{n}f_{n}(y).
\end{align}
Inserting it into ${\cal E}[\Phi]$ leads to
\begin{align}
		{\cal E}[\Phi] =\sum^{\infty}_{n=0}m_{n}^2 |\phi_{n}|^2 +({\rm quartic\hspace{0.5em} terms\hspace{0.5em} in \hspace{0.5em}}\phi_{n} ),
\end{align}
where
\begin{align}
		m_{n}^2\equiv M^2 +E_{n}\hspace{0.3em},\hspace{1.3em}n=0,1,2,\cdots.
\end{align}
Note that the quartic terms are non-negative for any configurations 
of $\phi_{n}$ because they come from the term 
$\int dy \frac{\lambda}{4}|\Phi|^4 \geq 0$. 
It immediately follows that the vacuum configuration is given 
by $\langle \Phi \rangle =0$  (or $\langle \phi_{n}\rangle =0$ 
for any $n$) if $m_{n}^2 \geq 0$ for any $n$.
Actually, we are interested in the inverse of the above statement 
that $\Phi$ (or $\phi_{0}$) acquires a nontrivial VEV if $m_{0}^2 <0$ 
for the lowest eigenvalue $E_{0}$.
This implies that the gauge symmetry breaking can occur even for
$M^{2}>0$ if a bound state exists in the spectrum.
On the other hand, the gauge symmetry can still be unbroken 
even for $M^{2}<0$ if there is no bound state with $E_{0}>0$.
Therefore, in order to determine whether or not the gauge symmetry
breaking occurs, we need to solve the bound state problem
of Eq.(\ref{eq:3-12}) with the BC's (\ref{eq:3-13}) and find 
the lowest energy eigenvalue $E_{0}$.

We can assume any bound state solution $f_{E<0}(y)$ and
any positive energy solution $f_{E>0}(y)$, without loss of
generality, to be of the form\footnote{
Zero energy solutions are given by $f_{E=0}(y)=a+by$.
}
\begin{align}
f_{E<0}(y) 
 &= a\, e^{\kappa(y-L/2)} + b\, e^{-\kappa(y-L/2)} \quad
   \textrm{with}\ E=-\kappa^{2}<0, \label{eq:3-17}\\
f_{E>0}(y) 
 &= A\, e^{ik(y-L/2)} + A^{*} e^{-ik(y-L/2)} \quad
   \textrm{with}\ E=k^{2}>0, \label{eq:3-18}   
\end{align}
where $a, b, \kappa$ and $k$ are real numbers with
$\kappa, k >0$ and $A$ is a complex one.
Inserting the above expressions into the BC's (\ref{eq:3-13})
and requiring nontrivial solutions to exist, we find the
equations to determine the energy spectrum, i.e.
\begin{align}
\tanh (\kappa L)
 &= \frac{\kappa(L_{+}+L_{-})}{1+\kappa^{2}L_{+}L_{-}},
  \label{eq:3-19}\\
\tan (k L)
 &= \frac{k(L_{+}+L_{-})}{1-k^{2}L_{+}L_{-}}.
  \label{eq:3-20}   
\end{align}
As was mentioned above, the criterion of the gauge symmetry
breaking is
\begin{align}
m_{0}^{\,2} = M^{2} + E_{0} < 0
  \label{eq:3-21}   
\end{align}
for the lowest eigenvalue $E_{0}$.
Noting that the parameters $L_{\pm}$ appear only in the
symmetric combinations $L_{+}L_{-}$ and $L_{+}+L_{-}$ in the
transcendental equations (\ref{eq:3-19}) and (\ref{eq:3-20}),
we find four distinct patterns of the spectrum, according
to the signs of $L_{+}L_{-}$ and $L_{+}+L_{-}$.
For the following discussions, it is convenient to introduce
the maximum and minimum values of the set $\{L_{+}, L_{-}\}$
\begin{align}
L_{\textrm{max}} \equiv \textrm{max}\{L_{+}, L_{-}\},\quad
L_{\textrm{min}} \equiv \textrm{min}\{L_{+}, L_{-}\}.
  \label{eq:3-22}   
\end{align}

%
%
%
\begin{itemize}
%
\item[(a)] $L_{+}L_{-}>0$ and $L_{+}+L_{-}>0$

Let us first consider the case of $L_{\pm}>0$, which may
be interpreted as the presence of two attractive 
$\delta$-function potentials at the boundaries.
In this case, there exist two bound states for
$L>L_{+}+L_{-}$ and a single one for $L\le L_{+}+L_{-}$.
The lowest eigenvalue is found to satisfy 
$E_{0}<-1/(L_{\textrm{min}})^{2}$ (see Fig.1(a)).
It follows that the gauge symmetry is spontaneously broken
for $M^{2}<1/(L_{\textrm{min}})^{2}$, because 
$m_{0}^{\,2}=M^{2}+E_{0}<0$.
For $M^{2}>1/(L_{\textrm{min}})^{2}$, there exists a 
critical length $L_{\textrm{c}}$ defined by
\begin{align}
L_{\textrm{c}}
 = \frac{1}{|M|} \textrm{arctanh} \biggl(
    \frac{|M|(L_{+}+L_{-})}{1+M^{2}L_{+}L_{-}}\biggr)
    \quad \textrm{for}\ M^{2}>\frac{1}{(L_{\textrm{min}})^{2}},
  \label{eq:3-23}   
\end{align}
and the gauge symmetry is broken (unbroken) for
$L<L_{\textrm{c}}$ ($L\ge L_{\textrm{c}}$).
The phase diagram is schematically depicted in Fig.2(a).
%
%
%
\begin{figure}[htbp]
		\begin{minipage}{0.47\textwidth}
		\begin{center}
		\scalebox{0.8}{\includegraphics{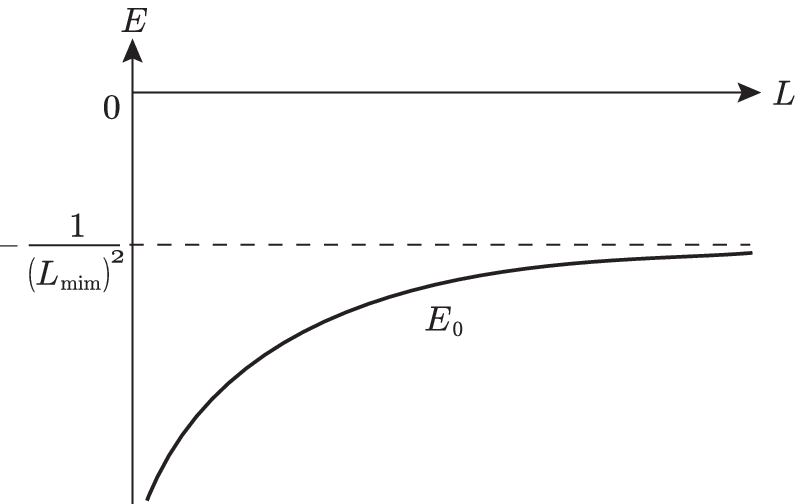}}
		\par
		\vspace{0.0cm}
		{\footnotesize{(a) $L_{+}L_{-}>0$, $L_{+}+L_{-}>0$}}
		\end{center}
		\end{minipage}
		\hfill
		\begin{minipage}{0.47\textwidth}
		\begin{center}
		\scalebox{0.8}{\includegraphics{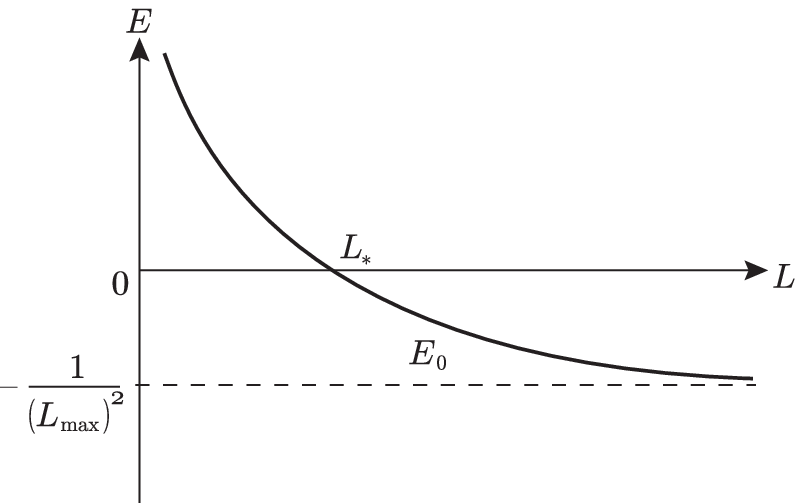}}
		\par
		\vspace{0.0cm}
		{\footnotesize{(b) $L_{+}L_{-}\leq0$, $L_{+}+L_{-}>0$}}
		\end{center}
		\end{minipage}
		\begin{minipage}{0.47\textwidth}
		\begin{center}
		\scalebox{0.8}{\includegraphics{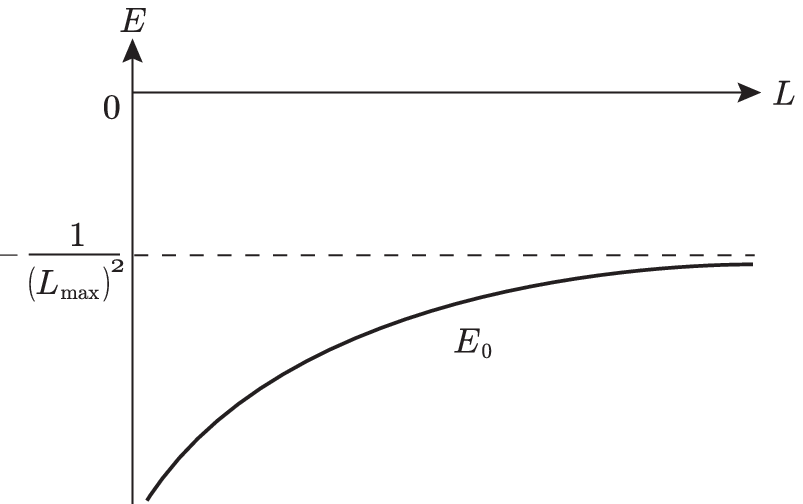}}
		\par
		\vspace{0.0cm}
		{\footnotesize{(c) $L_{+}L_{-}<0$, $L_{+}+L_{-} \le 0$}}
		\end{center}
		\end{minipage}
		\hfill
		\begin{minipage}{0.47\textwidth}
		\begin{center}
		\scalebox{0.8}{\includegraphics{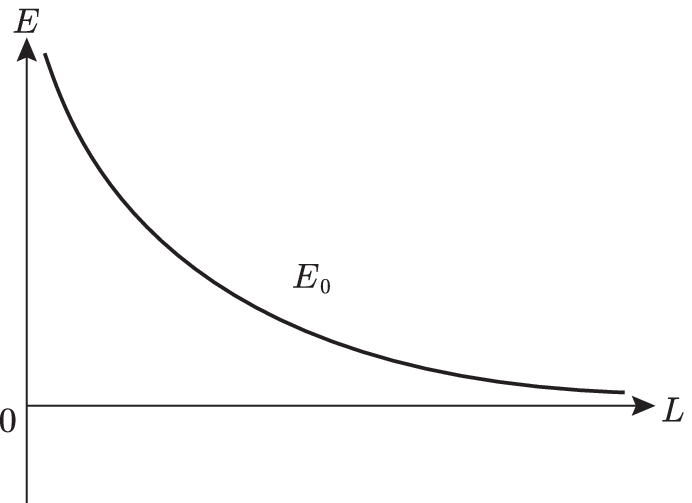}}
		\par
		\vspace{0.0cm}
		{\footnotesize{(d) $L_{+}L_{-}\ge 0$, $L_{+}+L_{-}\le 0$}}
		\end{center}
		\end{minipage}
		\vspace{0.05cm}\\
		\caption{The lowest energy spectrum $E_{0}$.
		         $L_{*}$ is given by $L_{*}=L_{+}+L_{-}$.
		 }
\end{figure}
\begin{figure}[htbp]
		\begin{minipage}{0.47\textwidth}
		\begin{center}
		\scalebox{0.8}{\includegraphics{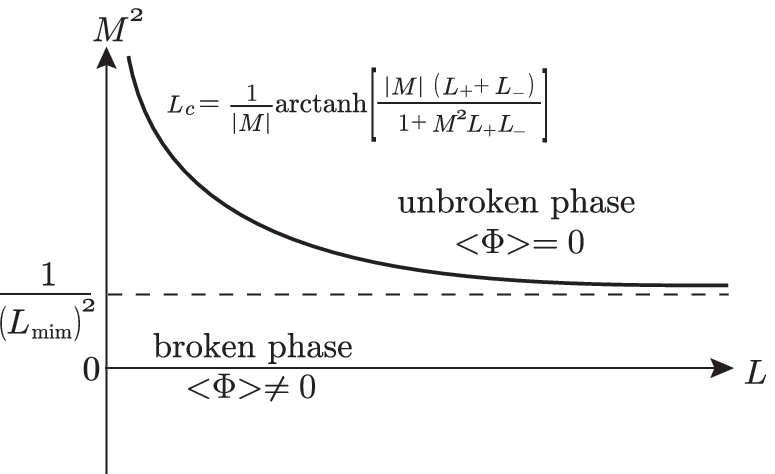}}
		\par
		\vspace{0.0cm}
		{\footnotesize{(a) $L_{+}L_{-}>0$, $L_{+}+L_{-}>0$}}
		\end{center}
		\end{minipage}
		\hfill
		\begin{minipage}{0.47\textwidth}
		\begin{center}
		\scalebox{0.8}{\includegraphics{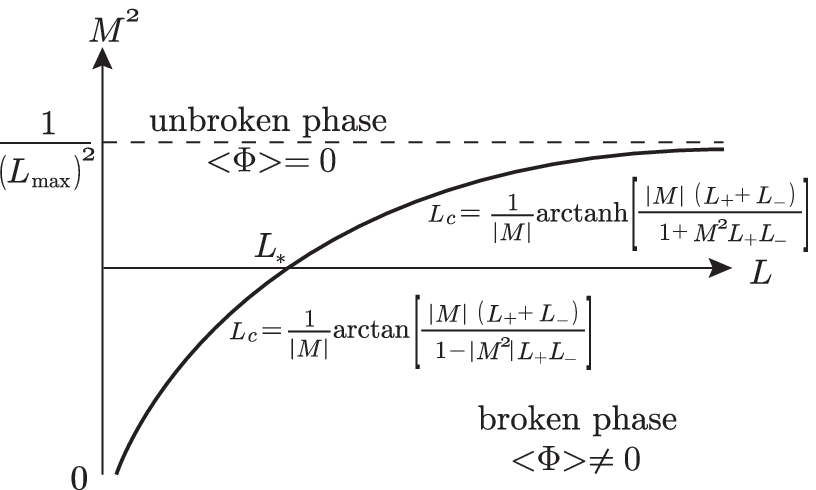}}
		\par
		\vspace{0.0cm}
		{\footnotesize{(b) $L_{+}L_{-}\le 0$, $L_{+}+L_{-}>0$}}
		\end{center}
		\end{minipage}
		\begin{minipage}{0.47\textwidth}
		\begin{center}
		\scalebox{0.8}{\includegraphics{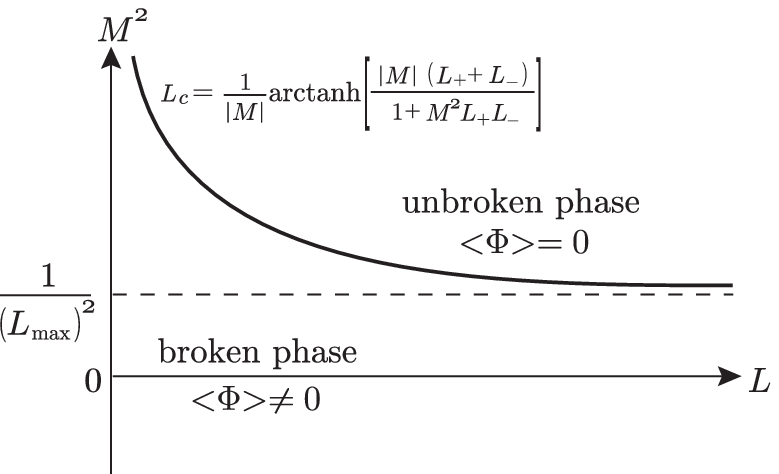}}
		\par
		\vspace{0.0cm}
		{\footnotesize{(c) $L_{+}L_{-}<0$, $L_{+}+L_{-} \le0$}}
		\end{center}
		\end{minipage}
		\hfill
		\begin{minipage}{0.47\textwidth}
		\begin{center}
		\scalebox{0.8}{\includegraphics{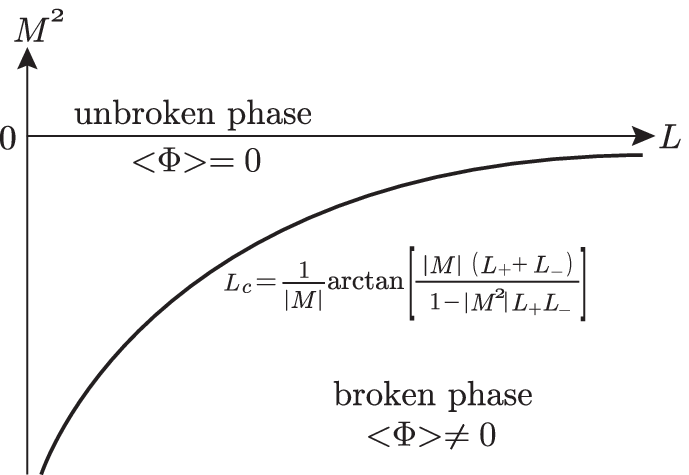}}
		\par
		\vspace{0.0cm}
		{\footnotesize{(d) $L_{+}L_{-}\ge 0$, $L_{+}+L_{-}\le 0$}}
		\end{center}
		\end{minipage}
		\vspace{0.05cm}\\
		\caption{Phase diagrams on an interval.
		 }
\end{figure}
%
%
%

\vspace{2mm}
%
\item[(b)] $L_{+}L_{-}\le 0$ and $L_{+}+L_{-}>0$

Let us next consider the case of $L_{\textrm{max}}>0$
and $L_{\textrm{min}}\le 0$ with $L_{\textrm{max}} > |L_{\textrm{min}}|$,
which may be interpreted as the presence of a relatively weak attractive
$\delta$-function potential and a relatively strong repulsive
$\delta$-function potential at the boundaries.
In this case, there is a bound state for $L>L_{*}\equiv L_{+}+L_{-}$
with $0 > E_{0} > -1/(L_{\textrm{max}})^{2}$.
For $L<L_{*}$, there is no bound state and the lowest energy
$E_{0}$ is positive (see Fig.1(b)).
Thus, for $M^{2}\ge 1/(L_{\textrm{max}})^{2}$, $m_{0}^{\,2}$
is always non-negative and hence the gauge symmetry is unbroken.
For $M^{2}< 1/(L_{\textrm{max}})^{2}$, the gauge symmetry is
broken (unbroken) for $L>L_{\textrm{c}}$ ($L\le L_{\textrm{c}}$),
where the critical length $L_{\textrm{c}}$ is defined by
\begin{align}
L_{\textrm{c}}
 = \left\{
    \begin{array}{ll}
      \frac{1}{|M|} \textrm{arctanh} \biggl(
        \frac{|M|(L_{+}+L_{-})}{1+M^{2}L_{+}L_{-}}\biggr)
         & \textrm{for}\ 0 < M^{2} < \frac{1}{(L_{\textrm{max}})^{2}},\\
      \frac{1}{|M|} \textrm{arctan} \biggl(
        \frac{|M|(L_{+}+L_{-})}{1-|M|^{2}L_{+}L_{-}}\biggr)
         & \textrm{for}\ M^{2}< 0,
    \end{array}  \right.
  \label{eq:3-24}   
\end{align}
Here, the values of $\textrm{arctan}(x)$ should be chosen
in the range of $0< \textrm{arctan}(x) < \pi$.
The resulting phase diagram is schematically depicted in
Fig.2(b).

\vspace{2mm}
%
\item[(c)] $L_{+}L_{-}<0$ and $L_{+}+L_{-}\le 0$

Let us next consider the case of $L_{\textrm{max}}>0$ and
$L_{\textrm{min}}<0$ with $L_{\textrm{max}} \le |L_{\textrm{min}}|$,
which may be interpreted as the presence of a relatively strong
attractive $\delta$-function and a relatively weak repulsive
$\delta$-function potential at the boundaries.
In this case, there is a single bound state in the spectrum
with $E_{0} < -1/(L_{\textrm{max}})^{2}$ (see Fig.1(c)).\footnote{
For the case of $L_{+}+L_{-}=0$, $E_{0}$ is given by
$E_{0}=-1/(L_{\textrm{max}})^{2}$, which may be obtained 
by taking the limit of $L_{+}+L_{-} \to -0$.
}
Thus, for $M^{2}<1/(L_{\textrm{max}})^{2}$, $m_{0}^{\,2}$ is always
negative, and hence the gauge symmetry is broken.
For $M^{2}>1/(L_{\textrm{max}})^{2}$, the gauge symmetry is broken
(unbroken) for $L<L_{\textrm{c}}$ ($L\ge L_{\textrm{c}}$),
where the critical length is defined by
\begin{align}
L_{\textrm{c}}
 = \frac{1}{|M|} \textrm{arctanh} \biggl(
    \frac{|M|(L_{+}+L_{-})}{1+M^{2}L_{+}L_{-}}\biggr)
    \quad \textrm{for}\ M^{2}>\frac{1}{(L_{\textrm{max}})^{2}}.
  \label{eq:3-25}   
\end{align}
The resulting phase diagram is schematically depicted in
Fig.2(c).\footnote{
For $L_{+}+L_{-}=0$, the critical line is given by
$M^{2}=1/(L_{\textrm{max}})^{2}$.
}

\vspace{2mm}
%
\item[(d)] $L_{+}L_{-}\ge 0$ and $L_{+}+L_{-}\le 0$

Let us finally consider the case of $L_{\pm}\le 0$,
which may be interpreted as the presence of two repulsive
$\delta$-function potentials at the boundaries.
In this case, there is no bound state and $E_{0} > 0$
(see Fig.1(d)).
Thus, for $M^{2}>0$, $m_{0}^{\,2}$ is always positive and 
hence the gauge symmetry is unbroken.
An interesting observation is that even if $M^{2}$ is
negative, the gauge symmetry is unbroken for $L>L_{\textrm{c}}$,
where the critical length is defined by
\begin{align}
L_{\textrm{c}}
 = \frac{1}{|M|} \textrm{arctan} \biggl(
    \frac{|M|(L_{+}+L_{-})}{1-|M|^{2}L_{+}L_{-}}\biggr)
    \quad \textrm{for}\ M^{2}<0.
  \label{eq:3-26}   
\end{align}
Here, the values of $\textrm{arctan}(x)$ should be chosen
in the range of $0< \textrm{arctan}(x) < \pi$.
The resulting phase diagram is schematically depicted in 
Fig.2(d).

\end{itemize}
%
%
%

\vspace{2mm}
In the above analysis, we have succeeded to determine the 
broken/unbroken phases of the gauge symmetry in the parameter 
space of the theory.
To derive them, we did not need the exact value of 
$\langle \Phi(y)\rangle$ but it was sufficient to know whether 
or not $\langle \Phi(y)\rangle$ is nonvanishing.
However, the exact value will be required to obtain the gauge boson 
mass in the broken phase and masses of the Kaluza-Klein modes 
of the scalar field.
According to similar analyses given in 
Refs. \cite{MS1988,LMT1992,STT1999a} with the different BC's, 
we can show that the exact VEV of $\langle\Phi(y)\rangle$ is given 
in terms of Jacobi elliptic functions and that the phase diagrams 
are precisely reproduced.
The Kaluza-Klein mass spectrum of the scalar field turns out to 
be governed by Lam\'e-type equations.
The details will be reported in a separate paper \cite{FNNOS2011}.

\section{Fermion BC's and 4d Chiral Zero Mode}
%
%
In the previous section, we have succeeded to reveal a rich phase 
structure of the scalar QED on the interval.
We would like to extend our analysis to gauge theories coupled to fermions.
To this end, we add the fermionic terms
\begin{align}
	S_{F}=\int d^4 x\int^{L}_{0}dy\hspace{0.3em} \bar{\Psi}(iD_{\mu}\gamma^{\mu}+iD_{y}\gamma^{y}+M_{F})\Psi\label{eq:4-1}
\end{align}
to the action (\ref{eq:3-1}).
Here, $\Psi(x,y)$ is a 4-component Dirac spinor and 
\begin{align}
	D_{N}\Psi =(\partial_{N}-ie_{F}A_{N})\Psi.
\end{align}
Note that the bulk fermion mass $M_{F}$ should be included 
in Eq.(\ref{eq:4-1}) because there is no Weyl fermion in 5-dimensions.
The extra component $\gamma^{y}$ of the gamma matrices can be chosen as
\begin{align}
	\gamma^{y}=-i\gamma^{5}=\gamma^{0}\gamma^{1}\gamma^{2}\gamma^{3}.
\end{align}

To obtain consistent BC's for $\Psi$, we require the action principle\!\!
\footnote{
The other requirements, such as the hermiticity of the action 
and the fermion number conservation, will lead to the same conclusion.
}
and find that $\bar{\Psi}\gamma^{y}\delta\Psi =0$ at $y=0,L$.
The condition may be decomposed into 
$\bar{\Psi}_{+}\delta\Psi_{-}=\bar{\Psi}_{-}\delta\Psi_{+}=0$ 
at $y=0,L$, where $\Psi_{\pm}$ are chiral spinors defined by 
$\gamma^{5}\Psi_{\pm}=\pm\Psi_{\pm}$.
It follows that the consistent BC's for a fermion are found to be
\begin{align}
	\Psi_{+}=0\hspace{.8em}{\rm or}\hspace{0.8em}\Psi_{-}=0 \hspace{1.7em}{\rm at}\hspace{0.5em} y=0,L.
\end{align}
The action principle also leads to the bulk equation, i.e. the 5d Dirac equation $(iD_{\mu}\gamma^{\mu}+D_{y}\gamma^{5}+M_{F})\Psi=0$.
In terms of $\Psi_{\pm}$, the Dirac equation is decomposed as 
\begin{align}
	iD_{\mu}\gamma^{\mu}\Psi_{+} +(-D_{y}+M_{F})\Psi_{-}=0,\\
	iD_{\mu}\gamma^{\mu}\Psi_{-} +(D_{y}+M_{F})\Psi_{+}=0.
\end{align}
The above equations imply that $\Psi_{+}=0$ 
($\Psi_{-}=0$) at $y=0$ or $L$ automatically gives the BC for 
$\Psi_{-}$ ($\Psi_{+}$) as $(-D_{y}+M_{F})\Psi_{-}=0$ 
$((D_{y}+M_{F})\Psi_{+}=0)$ at $y=0$ or $L$.
	We thus conclude that the fermion should obey one of 
the following four BC's:
\begin{align}
	&{\rm type}(++):(D_{y}+M_{F})\Psi_{+}=\Psi_{-}=0 \hspace{2.5em}{\rm at}\hspace{0.5em} y=0\hspace{0.3em}{\rm and}\hspace{0.3em}L,\nonumber\\
	&{\rm type}(--):\Psi_{+}=(-D_{y}+M_{F})\Psi_{-}=0 \hspace{1.7em}{\rm at}\hspace{0.5em} y=0\hspace{0.3em}{\rm and}\hspace{0.3em}L,\nonumber\\
	&{\rm type}(+-):(D_{y}+M_{F})\Psi_{+}=\Psi_{-}=0 \hspace{2.5em}{\rm at}\hspace{0.5em} y=0,\nonumber\\
	&\hspace{5em}\Psi_{+}=(-D_{y}+M_{F})\Psi_{-}=0 \hspace{1.7em}{\rm at}\hspace{0.5em} y=L,\nonumber\\
	&{\rm type}(-+):\Psi_{+}=(-D_{y}+M_{F})\Psi_{-}=0 \hspace{1.7em}{\rm at}\hspace{0.5em} y=0,\nonumber\\
	&\hspace{5em}(D_{y}+M_{F})\Psi_{+}=\Psi_{-}=0 \hspace{2.5em}{\rm at}\hspace{0.5em} y=L.
\end{align}
This result is very important in constructing phenomenological models 
on an interval because a chiral 4d fermion $\psi_{+}$ ($\psi_{-}$) 
appears in the 4d spectrum if a 5d fermion obeys the type($++$) 
(type($--$)) BC, while all the fermions with type($\pm\mp$) BC's 
will be decoupled from the low energy spectrum.
It is important to note that chiral 4d fermions are exponentially 
localized at boundaries.
With the type($++$) BC, every 4d chiral zero mode is localized 
at $y=0$ for $M_{F}>0$ ($y=L$ for $M_{F}<0$) according to 
the profile $\sim e^{-M_{F}y}$.
On the other hand, with the type($--$) BC, every 4d chiral zero mode 
is localized at $y=L$ for $M_{F}>0$ ($y=0$ for $M_{F}<0$) according 
to the profile $\sim e^{M_{F}y}$.
It should be emphasized that the bulk mass $M_{F}$ has nothing
to do with the presence or absence of a chiral zero mode but affects
its profile, and also that the analysis with the introduction
of Yukawa interactions will be  performed in a similar way.

To get a phenomenological model, we need to extend our analysis 
to nonabelian gauge theories.
Our mechanism to break gauge symmetries still works for them.
A simple extension of the SM, as a starting point to construct 
a realistic model beyond the SM, may be given as follows.
The 4d gauge fields of the SM are replaced by the 5d gauge fields 
with the BC's similar to Eqs.(\ref{eq:3-4}) and (\ref{eq:3-5}),\!\!\!\!
\footnote{
We will have a more variety of BC's than those considered in this paper 
in nonabelian gauge theories with many flavors.
}
which are consistent with 4d gauge symmetries of the SM.\!\!\!
\footnote{
Since there are no massless zero modes of the extra components $A_{y}$ 
of the gauge fields with the BC (\ref{eq:3-5}), the Hosotani 
mechanism \cite{Manton1979,Fairlie1979,Hosotani1983&89} to break 
gauge symmetry does not work in this model.
}	
The 4d chiral fermions of the SM have to be replaced by 5d Dirac 
fermions with their bulk masses.
Assuming that the 5d fermions have the same quantum numbers as 
the SM fermions, we impose the type($++$) BC for the $SU(2)$ singlet 
fermions and the type($--$) BC for the $SU(2)$ doublet ones.
Then, we may have desired 4d chiral fermions of the SM 
at low energies irrespective of the bulk fermion masses.
A key ingredient of our model is the choice of nontrivial BC's 
(\ref{eq:3-3}) for the Higgs field, which generate a nontrivial 
$y$-dependent VEV $\langle\Phi(y)\rangle$ and the electroweak 
gauge symmetry breaking.
Since bulk fermion masses do not provide masses of 4d chiral
fermions, as was noted above, they should acquire their masses 
through Yukawa interactions with localized profiles
of chiral zero modes and the Higgs VEV $\langle\Phi(y)\rangle$.
Thus, we expect the model to mimic the SM at low energies as a simple 
realization of the scenario given by \cite{DS2000,KT2001}.\!\!\!\!
\footnote{
A challenging attempt may be to introduce many branes or point 
singularities \cite{HSST1999,NST2003&04&05} on an interval, 
in which several copies of chiral fermions will appear.
Then, we can attack the generation problem of the SM together 
with the fermion mass hierarchy one.
}
The work along this line will be reported elsewhere \cite{FNNOS2011}.

\section{Conclusions}
%
We have investigated the nature of gauge symmetry breaking 
in gauge theories coupled with a scalar field on an interval.
We first derived the consistent set of boundary conditions 
for a scalar field.
These scalar BC's are characterized by two real parameters.
We have checked that they are compatible with the various 
consistency requirements; the action principle, the gauge 
invariance, the hermiticity of the action and the charge conservation.
Allowing general BC's for the scalar field, we have observed 
that the scalar can develop a nonvanishing VEV to break gauge symmetry, 
like the Higgs field of the SM.
The gauge symmetry breaking mechanism is, however, quite different 
from the usual Higgs one.
We do not need a negative mass square term to break gauge symmetry.
The scalar field can acquire a nontrivial VEV even if its mass square 
is positive.
Any nonvanishing value of the scalar field cannot be a constant 
but inevitably depends on the extra dimensional coordinate.
The phase diagram is found to depend nontrivially on the length of 
the interval, the mass and the BC's of the scalar field.

Since the main purpose of this paper is to demonstrate our gauge 
symmetry breaking mechanism, we have restricted our considerations 
to a simple $U(1)$ gauge theory.
However, the extension to nonabelian gauge theories is straightforward.
We will then have a large variety of consistent BC's to break gauge 
symmetries.
As was discussed in Section 4, 4d chiral fermions naturally arise 
from bulk fermions  on an interval even if their bulk masses are 
nonvanishing, and they are, in general, localized at one of the boundaries.
This is good news to solve the fermion mass hierarchy problem.
Chiral fermions will acquire their masses through Yukawa couplings.
We may then have a chance to get the hierarchical fermion masses of 
the SM with the localization property of chiral zero modes together 
with the extra dimensional coordinate dependence of the VEV 
$\langle\Phi(y)\rangle$.

In this paper, we have discussed the phase structure of gauge
theories on an interval at the tree level.
Quantum effects may, however, change our results because they
will produce mass corrections to the Higgs scalar, which, in general,
depend on the scale of the extra dimension.
Such radiative corrections would become important when a 
compactification  scale becomes less than the inverse of a typical
mass scale of the theory, and then some of broken symmetries could
be restored or conversely some of symmetries could be broken, as
shown in Ref. \cite{HMOS2001}.
Thus, our analyses at the tree level will be insufficient and
it would be of great importance to investigate quantum corrections
in a class of theories we considered.

\section*{Acknowledgements}
	This work is supported in part by a Grant-in-Aid for Scientific Research (No.22540281 and No.20540274 (M.S.)) from the Japanese Ministry of Education, Science, Sports and Culture.
	The authors would like to thank N.Maru, K.Nishiwaki, N.Sakai and K.Takenaga for valuable discussions.

%

\end{document}